# Field-induced inhomogeneous ground states of antiferromagnetic ANNNI chains.

*P.N. Timonin*

*Southern Federal University, 344090, Rostov-on-Don, Russia*

*Finite-size effects are studied in ground states of antiferromagnetic (AF) ANNNI chains in a field. It is shown that field can induce a variety of inhomogeneous states in finite chains. They are composed of two shifted AF states with the "kink" at their junction and are highly degenerate with respect to the kink position. The phase diagram 'field – exchange ratio' for finite chains is presented.*

## 1. Introduction

Finite-size effects in the antiferromagnetic (AF) ground states of short-range Ising chains in a field have some definite peculiarity: the possible perturbations of AF spin order caused by free boundaries cannot be confined to some vicinity of chain ends but spread throughout the whole chain. This is the consequence of the AF ground states' degeneracy in infinite chains (or rings) and, hence, of the infinite correlation length. The origin of degeneracy lies in the translational invariance of infinite systems and rings, generally there can be several ground states shifted along the chain.

Consider the simplest case of finite chain with nearest-neighbor (NN) AF exchange in a field $H$. If it has odd number of spins $N$ free boundaries lift the degeneracy making the state with end spins up (along the field) the unique ground state at $0 < H < 2J$, $J$ is AF exchange. This apparently changes spin order throughout the whole chain. Yet more pronounced effect appears in case of even number of spins: while at $0 < H < J$ there is still usual two-fold degeneracy, at $J < H < 2J$ $N/2$ degenerate ground states appear. Its construction consists in division of chain in two segments with odd numbers of spins and endowing each segment with the spin structure of odd chain, i.e. end spins up in both segments, see Fig. 1a. Thus each such state has a "kink" composed of two nearby spins up with the usual AF order beyond the kink. It can be readily verified that it has lower energy than perfect AF order at $J < H < 2J$. Such kink states were first found in macroscopic model of magnetic layers with AF exchange [1]. In the ensemble of nearest-neighbor AF chains the averaging over these states gives the following linearly-modulated AF order for average Ising spins [2], see Fig. 1b,

$$m_n = \langle s_n \rangle = N^{-1}\left[1 + (-1)^n (N - 1 - 2n)\right], \quad n = 0, 1, ..., N-1. \quad (1)$$

Here number of spins $N$ is even. We may note that ensembles of such AF chains can be realized in dilute crystals with (quasi) $1d$ magnetic structure, in solutions of magnetic polymers and in magnetic liquid crystals. Intensity of magnetic neutron diffraction on these objects in such "bow-tie" phase bears only traces of the underlying AF order exhibiting a broad peak around $k = \pi$ [2].

Similar field-induced inhomogeneous ground states may exist in more complex Ising models on finite chains and chain-like structures (stripes, tubes) with interactions spreading over several unit cells. In such models it may not be easy to find ground states via enumeration of possible variants and comparing their energies. Here the regular calculations within transfer-matrix (TM) formalism can be used at low temperatures. In Section 2 we consider simple AF Ising chain with NN exchange to find its ground state within this standard approach. In Section 3 we describe TM formalism for the Ising chain with NN and next-nearest-neighbor (NNN) AF exchange (ANNNI model) and apply it in Sections 4, 5 to reveal the appearance field-induced inhomogeneous ground states in this model.

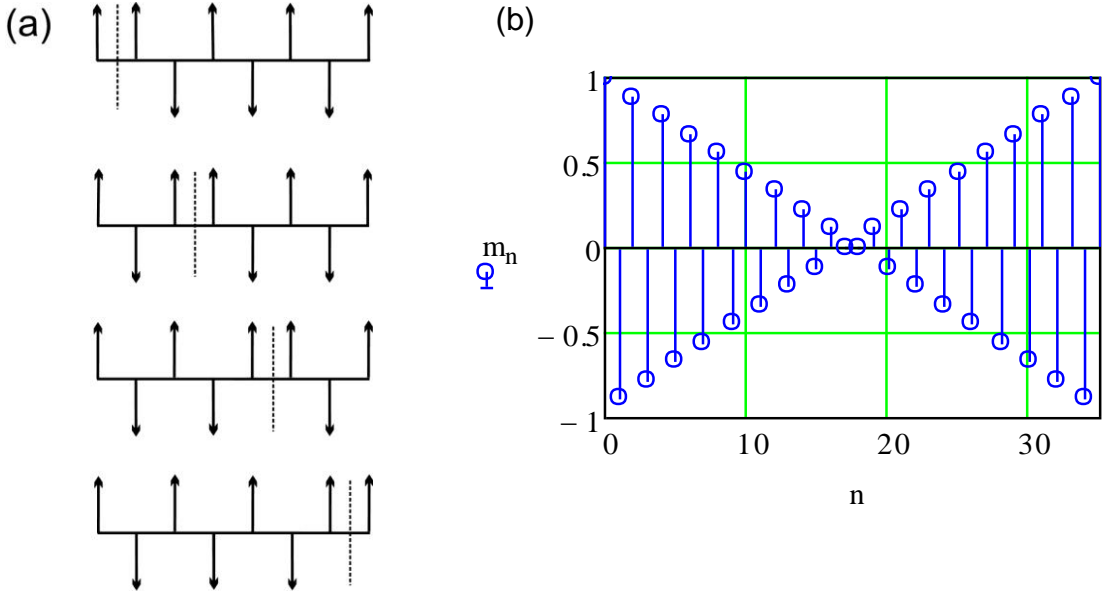

Fig.1. (a) Kink ground states of NN AF chain with $N = 8$ at $J < H < 2J$, dotted lines mark kink position; (b) resulting average magnetization profile for $N = 32$ from Eq. (1).

## 2. Ground states of NN AF Ising chain.

Let us see how TM formalism works for a simple (NN) AF chain and how Eq. (1) emerges in it. The Hamiltonian of this model

$$\mathcal{H} = J \sum_{n=0}^{N-2} s_n s_{n+1} - H \sum_{n=0}^{N-1} s_n$$

defines the TM

$$U_{s,s'} = \exp\left[-K(1+ss') + h(s+s')/2\right], \quad K=J/T,\ h=H/T, \qquad (2)$$

which is used to find average local spins as

$$m_n \equiv \langle s_n \rangle = Tr\left(\hat{V}\hat{U}^{n'}\hat{\sigma}_z\hat{U}^n\right) / Tr\left(\hat{V}\hat{U}^L\right), \quad L = N-1,\ n' = L - n \qquad (3)$$

Here $\hat{\sigma}_z$ is Pauli matrix,

$$V_{s,s'} = v_s v_{s'}, \quad v_s = \exp\left[h(s-1)/2\right]$$

for free boundaries and $\hat{V} = \hat{I}$ - identity matrix for a ring.

At low $T \ll \min(H,J)$ we have

$$\hat{U} = \begin{pmatrix} \kappa & 1 \\ 1 & 0 \end{pmatrix} = \hat{\sigma}_x + \kappa \begin{pmatrix} 1 & 0 \\ 0 & 0 \end{pmatrix}, \quad \kappa \equiv \exp(h-2K)$$

So at $H < 2J$ $\kappa \to 0$ for $T \to 0$ and eigenvalues of $\hat{U}$ are

$$\lambda_\mu \approx \mu + \kappa/2, \quad \mu = \pm 1.$$

Thus $\kappa$ defines the correlation length $\xi = 1/\ln|\lambda_{+1}/\lambda_{-1}| \approx \kappa^{-1}$ which diverges at $T \to 0$. Using spectral decomposition of $\hat{U}$ [3] we find

$$\hat{U}^L = \sum_{\mu=\pm 1} \lambda_\mu^L \hat{E}_\mu, \quad \hat{E}_\mu = (\hat{U} - \lambda_{-\mu}\hat{I})/(\lambda_\mu - \lambda_{-\mu}) \approx \frac{1}{2}\left[\hat{I} + \mu\left(\hat{\sigma}_x + \frac{\kappa}{2}\hat{\sigma}_z\right)\right]$$

At sufficiently low $T$ $\xi$ becomes much larger than $L$ ($\kappa L \ll 1$) and here we have

$$\hat{U}^L \approx |\lambda_{+1}\lambda_{-1}|^{L/2}\left[\hat{E}_{+1}\left(1 + \frac{\kappa L}{2}\right) + (-1)^L\left(1 - \frac{\kappa L}{2}\right)\hat{E}_{-1}\right] \approx$$

$$v_L^+\left(\hat{I} + \frac{\kappa L}{2}\hat{\sigma}_x\right) + v_L^-\left(\hat{\sigma}_x + \frac{\kappa}{2}\hat{\sigma}_z + \frac{\kappa L}{2}\hat{I}\right),$$

$$v_L^\pm \equiv \left[1 \pm (-1)^L\right]/2$$

Thus for $N$ even ($L$ odd) and free boundaries

$$Tr\hat{V}\hat{U}^L = 2e^{-h} + \frac{\kappa}{2}\left[1 - e^{-2h} + (1+e^{-2h})L\right] \approx \frac{\kappa}{2}\left(N + 4e^{2(K-h)}\right)$$

while

$$Tr\left(\hat{V}\hat{U}^{n'}\hat{\sigma}_z\hat{U}^n\right) = \frac{\kappa}{2}\left[1 + e^{-2h} + (-1)^n(N-1-2n)(1-e^{-2h})\right]$$

So at $0 < H < 2J$ for $N$ even we have approximately at $T \ll \min(H,J)$

$$m_n = \frac{1 + (-1)^n(N-1-2n)}{N + 4e^{2(K-h)}}$$

justifying the ground state expression (1) for $m_n$ at $J < H < 2J$.

At $0 < H < 2J$ free energy of even AF chains is (apart from irrelevant constant)

$$F = -\frac{T}{N}\ln\left(e^h Tr\hat{V}\hat{U}^L\right) \approx -\frac{T}{N}\ln\left(2 + \frac{e^{2(h-K)}N}{2}\right)$$

It describes the finite-size effects in the ensemble of even AF chains at low $T$.

In particular, the entropy

$$S = N^{-1}\left[\ln\left(2 + \frac{e^{2(h-K)}N}{2}\right) + \frac{2(K-h)N}{N + 4e^{2(K-h)}}\right]$$

experiences sharp growth from rather low values at $0 < H < J$

$$S \approx N^{-1}\ln 2 + e^{2(h-K)}(K-h)/2$$

to greater ones at $J < H < 2J$

$$S \approx N^{-1}\ln(N/2)$$

reflecting the appearance of $N/2$ ground states. Also magnetization

$$M = N^{-1}\sum_{n=0}^{N-1} m_n = -\frac{\partial F}{\partial H} = \frac{2}{N + 4e^{2(K-h)}}$$

grows from nearly zero to the constant value $2/N$ when $H$ becomes greater than $J$.

## 2. AF ANNNI chain

We consider finite Ising chain with first and second neighbor AF exchange having the Hamiltonian

$$\mathcal{H} = J_1\sum_{n=0}^{N-2} s_n s_{n+1} + J_2\sum_{n=0}^{N-3} s_n s_{n+2} - H\sum_{n=0}^{N-1} s_n$$

The ground states of this model on the infinite ring or chain are found in Ref. [4] as function of exchange ratio $\alpha = J_2/J_1$ and $H > 0$. There are four types of order: ordinary AF1 with (up, down) unit cell, AF2 with (up, up, down, down) cell, ferrimagnetic (up, up, down) and ferromagnetic one [4]. The TM of the model connects the neighboring pairs of spins

$$U_{\mathbf{s},\mathbf{s}'} = \exp\left\{-K_1\left[\alpha(\mathbf{s}\mathbf{s}' + 2) + s_2 s_1' + (s_1 s_2 + s_1' s_2')/2\right] + h(s_1 + s_2 + s_1' + s_2')/2\right\}$$

and local magnetization is

$$m_n = Tr\left(\hat{V}\hat{U}^{l'}\hat{S}_n\hat{U}^l\right)/Tr\left(\hat{V}\hat{U}^L\right), \quad L = \left[\frac{N}{2}\right] - 1, \quad l = \left[\frac{n}{2}\right] \quad l' = L - l, \tag{4}$$

$$\left(\hat{S}_n\right)_{s,s'} = \delta_{s,s'}\left(s_2 v_n^+ + s_1 v_n^-\right) = \delta_{s,s'}\left[s_1 + s_2 + (-1)^{n+1}(s_1 - s_2)\right]/2 \tag{5}$$

Square brackets in (4) denote the integer part of a number.

The chain with free boundaries and even $N$ has

$$V_{s,s'} = v_s v_{s'}, \quad v_s = \exp\frac{1}{2}\left[K_1(1-s_1 s_2) + h(s_1 + s_2)\right] \tag{6}$$

while for odd $N$

$$V_{s,s'} = v_s \tilde{v}_{s'}, \quad \tilde{v}_{s'} = v_{s'}\sum_{s=\pm 1}\exp s\left[-K_1(s_1' + \alpha s_2') + h\right] \tag{7}$$

Adopting the following convention for the relation between row (column) numbers and configurations $(s_1, s_2)$

$1-(++), 2-(--), 3-(+-), 4-(-+)$

we can represent $\hat{U}$ at $T \ll \min(J_1, J_2, H)$ as

$$\hat{U} = \begin{pmatrix} c^2 & 1 & c & a \\ 1 & 0 & b & 0 \\ a & 0 & ab & 1 \\ c & b & 1 & ab \end{pmatrix} \tag{8}$$

$a = \exp(h + K_1 - 2K_2), \quad b = \exp(-h + K_1 - 2K_2), \quad c = \exp(h - K_1 - 2K_2)$

Apparently, at low $T$ considered

$$a \gg \max(b, c), \quad bc \ll 1 \tag{9}$$

so the approximate equation for eigenvalues $\lambda$ of $\hat{U}$ reads

$$\left(\lambda^2 - ab\lambda - 1\right)^2 = \lambda(c\lambda + a)^2 \tag{10}$$

The solutions to Eq. (10) have four different types of behavior when $T \to 0$ depending on the relations between $a$, $b$ and $c$. For the eigenvalues with largest modules we have

$$b^{-3} \ll a \ (h_1 < 2 - 4\alpha), \quad \lambda_\mu \approx ab\left(1 + \frac{\mu}{\sqrt{ab^3}}\right), \quad \mu = \pm 1 \tag{11}$$

$$a \to 0 \ (h_1 < 2\alpha - 1), \ \lambda_{\mu,\tau} \approx \mu\left(1 + \tau\sqrt{\mu a}/2\right), \ \mu, \tau = \pm 1 \qquad (12)$$

$$\max\left(c^3, 1\right) \ll a \ll b^{-3} \ \left(3\left|\frac{1}{2} - \alpha\right| + \frac{1}{2} - \alpha < h_1 < 2 + 2\alpha\right),$$
$$\lambda_k \approx a^{2/3} e^{i2\pi k/3}, \ k = 0, 1, 2. \qquad (13)$$

$$a \ll c^3 \ (h_1 > 2 + 2\alpha), \ \lambda \approx c^2. \qquad (14)$$

Here $h_1 = H/J_1$.

So in four regions of $h_1 - \alpha$ plane $\hat{U}$ has different limiting forms at $T \to 0$, which we can obtain leaving in the original TM (8) only those matrix elements which give the eigenvalues listed above, namely,

$$h_1 < 2 - 4\alpha, \ \hat{U} \approx \begin{pmatrix} 0 & 0 & 0 & a \\ 0 & 0 & 0 & 0 \\ a & 0 & ab & 0 \\ 0 & 0 & 1 & ab \end{pmatrix} \qquad (15)$$

$$h_1 < 2\alpha - 1, \ \hat{U} \approx \begin{pmatrix} 0 & 1 & 0 & a \\ 1 & 0 & 0 & 0 \\ a & 0 & 0 & 1 \\ 0 & 0 & 1 & 0 \end{pmatrix} \qquad (16)$$

$$3\left|\frac{1}{2} - \alpha\right| + \frac{1}{2} - \alpha < h_1 < 2 + 2\alpha, \ \hat{U} \approx \begin{pmatrix} 0 & 0 & 0 & a \\ 0 & 0 & 0 & 0 \\ a & 0 & 0 & 0 \\ 0 & 0 & 1 & 0 \end{pmatrix} \qquad (17)$$

$$h_1 > 2 + 2\alpha, \ \hat{U} \approx \begin{pmatrix} c^2 & 0 & 0 & 0 \\ 0 & 0 & 0 & 0 \\ 0 & 0 & 0 & 0 \\ 0 & 0 & 0 & 0 \end{pmatrix} \qquad (18)$$

These different limiting forms of $\hat{U}$ make evident that we have different ground states in the corresponding regions of $h_1 - \alpha$ plane. Thus at $h_1 > 2\alpha - 1$ and $T \to 0$ $\hat{U}$ transfers $(++)$ into $(--)$ then to $(++)$ and so on, while $(+-)$ is transferred to $(-+)$ then to $(+-)$ etc. So here we have $(++--)$ phase AF2 in the infinite chain. At $h_1 < 2 - 4\alpha$ the sequence is $(+-) \to (+-) \to (+-)$ ... or $(-+) \to (-+) \to (-+)$ etc. and it is the ordinary AF1 phase.

At $3\left|\frac{1}{2}-\alpha\right|+\frac{1}{2}-\alpha < h_1 < 2+2\alpha$ spin structure has period 3 as matrix elements of $\hat{U}$ generate the sequence $(-+) \rightarrow (+-) \rightarrow (++)$, that is the ferrimagnetic ground state $(-++)$. At $h_1 > 2+2\alpha$ only $(++)$ state survives so this is the region of ferromagnetic ground state. This phase diagram found in Ref. [4] is shown in Fig. 2.

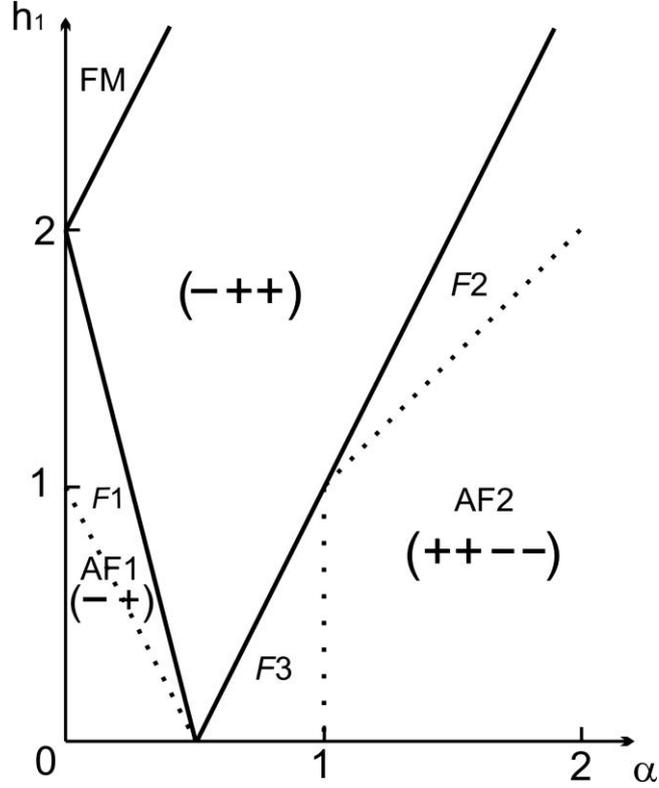

Fig.2. Ground state phase diagram for the finite AF ANNNI chain in a field. Solid lines are phase boundaries for infinite chains or rings found in Ref. [4]. Dotted lines show the boundaries of inhomogeneous (modulated) phases of finite chains.

In the ferrimagnetic phase the finite size effects are rather trivial consisting in (partial) lifting the triple degeneracy of the ground states. Nontrivial ground states in finite chains can appear in two AF phases of present model so further we consider just these phases.

### 4. Ground states in AF1 phase.

Here it is convenient to use the normalized TM $\hat{W} = (ab)^{-1}\hat{U}$ in which we omit the second row and second column with zeros, cf. (15),

$$\hat{W} = \begin{pmatrix} 0 & 0 & b^{-1} \\ b^{-1} & 1 & 0 \\ 0 & \varepsilon^2 b^2 & 1 \end{pmatrix}, \quad \varepsilon = \frac{1}{\sqrt{ab^3}}. \qquad (19)$$

At $T \rightarrow 0$ $\varepsilon$ goes to zero and for $\varepsilon L \ll 1$ we have from the spectral decomposition

$$Z_L = Tr(\hat{V}\hat{W}^L) \approx Tr(\hat{V}\hat{G}_+) + L\varepsilon Tr(\hat{V}\hat{G}_-), \quad \hat{G}_\pm = \hat{E}_+ \pm \hat{E}_-.$$

Here $\hat{E}_+$ and $\hat{E}_-$ are the idempotent components related to the eigenvalues $1+\varepsilon$ and $1-\varepsilon$ correspondingly. Apparently, $2\varepsilon$ is inverse correlation length. In the lowest order in $\varepsilon$ we have [3]

$$\hat{G}_+ = \begin{pmatrix} 0 & 0 & b^{-1} \\ b^{-1} & 1 & -b^{-2} \\ 0 & 0 & 1 \end{pmatrix}, \quad \varepsilon\hat{G}_- = \begin{pmatrix} 0 & 0 & 0 \\ 0 & 0 & b^{-2} \\ 0 & 0 & 0 \end{pmatrix}$$

so for $N$ even and $K_2 > 0$

$$Z_L = (2b^2 - 1 + L + 2e^{-2K_2})e^{2K_1}b^{-2} \approx (2b^2 - 1 + L)e^{2K_1}b^{-2}.$$

Similarly, we get

$$Z_L m_n \approx Tr(\hat{V}\hat{G}_+\hat{S}_n\hat{G}_+) + \varepsilon l Tr(\hat{V}\hat{G}_+\hat{S}_n\hat{G}_-) + \varepsilon l' Tr(\hat{V}\hat{G}_-\hat{S}_n\hat{G}_+)$$

Here $\hat{S}_n$ differs from that in Eq. (5) by the absence of second row and second column.

For $N$ even ($L$ odd) and $L \geq 3$, $l, l' \geq 1$, $K_2 > 0$ we have

$$m_n = \frac{1 + (-1)^n (l' - l)}{2b^2 - 1 + L}. \tag{20}$$

Thus for $b \to \infty$ ($h_1 < 1 - 2\alpha$) $m_n$ is zero as in ordinary doubly degenerate AF state but when $b \to 0$ ($1 - 2\alpha < h_1 < 2 - 4\alpha$, F1 region in Fig. 2) we have linearly modulated state. Returning to the site numbers in (20), cf. Eq. (4), we get

$$m_n = \frac{1 + (-1)^n (N - 1 - 2n)}{N - 4}, \quad N \geq 6, \quad 2 \leq n \leq N - 3$$

Fourier transform of this $m_n$,

$$\tilde{m}_k = \sum_{n=2}^{N-3} e^{ikn} m_n = \frac{e^{i3k/2}}{\cos(k/2)} + \delta(k, 0), \quad k = \frac{2\pi r}{N-4}, \quad r = 0, 1, ..., N-5,$$

depends on $N$ through $k$ values only, so it has definite thermodynamic limit and can be considered as order parameter for this inhomogeneous phase. Note that the same $\tilde{m}_k$ has the "bow-tie" phase of NN AF chain (1) for $k = 2\pi r / N$, cf. Ref. [2]. It defines the intensity of neutron diffraction $I_k \sim |\tilde{m}_k|^2$ for wave vector *k* expressed in units of inverse cell parameter.

The calculations for $l = 0$ and $l = L$ result in $m_0 = m_{N-1} = 1$, $m_1 = m_{N-2} = -1$ for $K_2 > 0$ and we have a profile shown in Fig. 3 for $N = 36$. Evidently, here we have $(N - 4)/2$ degenerate ground states formed from the inner $N - 4$ spins by the process similar to that in simple NN AF chain. The couples of the boundary spins stay fixed at the above values as NNN AF exchange forbids two parallel spins at the edges. For *N* odd $m_n = (-1)^n$ in all AF1 region.

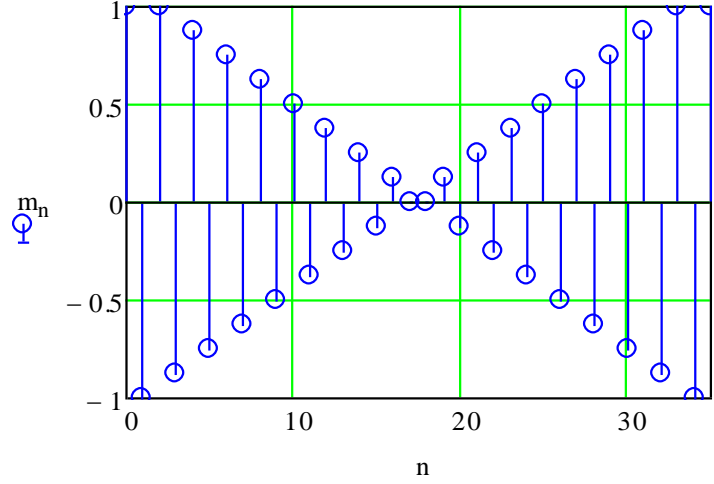

Fig.3. Magnetization of linearly modulated state of ANNNI chain with $N = 36$ at $1 - 2\alpha < h_1 < 2 - 4\alpha$ (*F1* region in Fig.2).

### 5. Ground states in AF2 phase.

According to (16) $\hat{U}$ in this phase ($h_1 < 2\alpha - 1$) can be represented as

$$\hat{U} = \hat{E} + a\hat{G}, \quad \hat{E} = \begin{pmatrix} \hat{\sigma}_x & 0 \\ 0 & \hat{\sigma}_x \end{pmatrix}, \quad \hat{G} = \begin{pmatrix} 0 & 0 & 0 & 1 \\ 0 & 0 & 0 & 0 \\ 1 & 0 & 0 & 0 \\ 0 & 0 & 0 & 0 \end{pmatrix}.$$

As $a \to 0$ for $T \to 0$, we can evaluate the powers of $\hat{U}$ without resorting to spectral decomposition. Thus at $ak \ll 1$

$$\hat{U}^{2k} \approx \hat{I} + ka\hat{D}, \quad \hat{D} = \{\hat{E}, \hat{G}\} = \begin{pmatrix} 0 & \hat{I} \\ \hat{\sigma}_x & 0 \end{pmatrix} \tag{21}$$

From this equation it follows that chains with *L* even do not have inhomogeneous ground states as in such case we can neglect the contributions of order *a* in $m_n$ calculations. However, for *L* odd (i. e. for $N = 4M$ or $N = 4M + 1$) we have

$$Z_L \approx Tr\hat{V}\hat{E} + \frac{a}{2}Tr\hat{V}\left[(L+1)\hat{G} + (L-1)\hat{E}\hat{G}\hat{E}\right] \qquad (22)$$

and $Tr\hat{V}\hat{E} = 0$ for $N = 4M$ while for $N = 4M + 1$ $Tr\hat{V}\hat{E}$ may go to zero as $T \to 0$, see (6, 7). Thus further we consider $L$ odd. In such a case

$$Z_L m_n \approx Tr\hat{V}\left[(-1)^{l'}\left(\hat{I} + a\left[\frac{l'}{2}\right]\hat{D}\right)\hat{E}\hat{S}_n\left(\hat{I} + a\left[\frac{l}{2}\right]\hat{D}\right) + av_l^+\hat{G}\hat{S}_n + av_l^-\hat{S}_n\hat{G}\right] \qquad (23)$$

For $N = 4M$ we get from these equations

$$m_n = \frac{(-1)^{l+n}(L+1) + 1 + (-1)^l(L-2l)}{2(L+1) + 4e^{2K_1(\alpha - h_1)}} \qquad (24)$$

Thus at $\alpha < h_1 < 2\alpha - 1$ (F2 region in Fig. 2) we have

$$m_n = N^{-1}\left\{1 + (-1)^l\left[v_n^+(N-1) - n\right]\right\} \qquad (25)$$

Fig. 4 shows $m_n$ (25) for $N = 64$. One can see that this is the result of averaging over $N/4 = M$ degenerate ground states which are formed from AF2 state via flipping odd number of successive spins in even or odd sublattice. They are shown in Fig. 5(b)-(e) for $N = 8$.

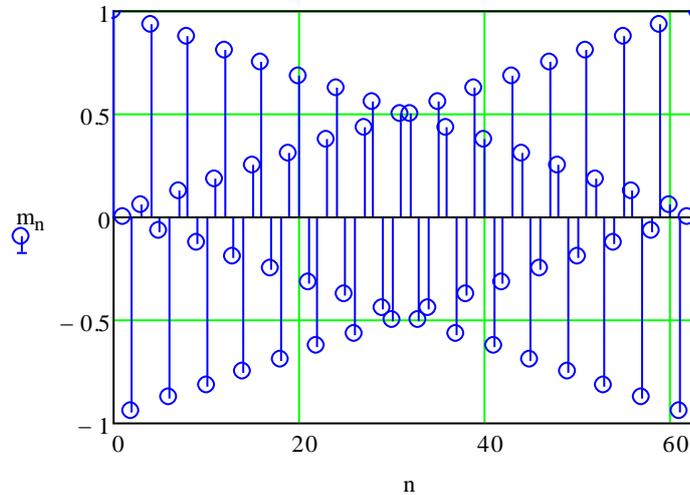

Fig.4. Magnetization profile at $\alpha < h_1 < 2\alpha - 1$ (F2 region in Fig. 2) for $N = 64$.

Fourier transform of (25) is

$$\tilde{m}_k = \sum_{n=0}^{N-1} e^{ikn} m_n = \frac{e^{-ik/2}\cos(k/2)}{\cos(k)}, \quad k = \frac{2\pi r}{N} = \frac{\pi r}{2M}, \quad r = 0, 1, ..., N-1,$$

Its singularities at $k = \pi/2$ and $k = 3\pi/2$ reflects the underlying nearly perfect period-4 order of the kink states.

According to (24) at $h_1 < \min(2\alpha-1, \alpha)$ $m_n \to 0$ when $T \to 0$ which means that chain is in doubly degenerate $(++--)$ AF2 ground states. For $N = 8$ one of these states is shown in Fig. 5 (a), its counterpart has all spin reversed.

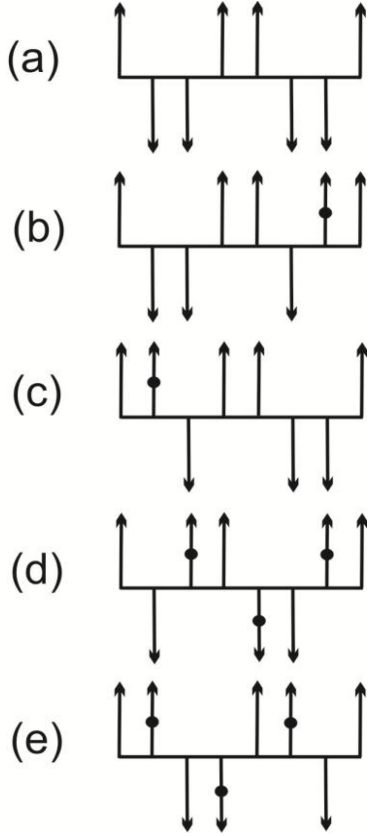

Fig.5. Procedure of ground states formation at $\alpha < h_1 < 2\alpha - 1$ for $N = 8$. (a) The parent state. The ground states are obtained via flipping in it one spin (b), (c) or three spins (d), (e). Flipped spins are marked by the dots.

For $N = 4M + 1$ and $|1-\alpha| < h_1 < \min(2\alpha-1, \alpha+1)$ we get from (7, 22, 23)

$$m_n = \frac{2v_n^-\left[1+(-1)^l\left(\left[\frac{l'}{2}\right]-\left[\frac{l}{2}\right]\right)\right]+v_n^+(-1)^l(L+1)+4v_n^+\left[e^{2K_1(1-h_1)}v_l^- + e^{2K_1(\alpha-h_1)}(-1)^l\right]}{(L+1)\left(1+2e^{2K_1(1-h_1)}\right)+4e^{2K_1(\alpha-h_1)}}$$

Calculating $m_n$ for other regions of $h_1 - \alpha$ plane where AF2 phase exists we finally get for $N = 4M + 1$:

$\alpha < h_1 < 2\alpha - 1$, F2 region in Fig. 2,

$$m_n = v_n^+(-1)^l + \frac{2v_n^-}{L+1}\left\{1+(-1)^l\left(\left[\frac{l'}{2}\right]-\left[\frac{l}{2}\right]\right)\right\}; \qquad (26)$$

$\alpha < 1$, $h_1 < 2\alpha - 1$, F3 region in Fig. 2, $m_n = \dfrac{2v_n^+ v_l^-}{L+1} = \dfrac{1}{M}\sum_{r=1}^{M}\delta(n, 4r-2);$ \qquad (27)

$\max(1, h_1) < \alpha$, the rest of AF2 phase in Fig. 2, $\quad m_n = v_n^+ (-1)^l.$ \quad (28)

The profile (26) for $\alpha < h_1 < 2\alpha - 1$ (F2 region) is shown in Fig. 6 for $N = 65$. We see that even sublattice have AF order with up spins at the edges but the odd one is formed via averaging over $(N-1)/4 = M$ kink states. In contrast, at $\max(1, h_1) < \alpha$ the AF order in odd sublattice stay intact up to a global reversal so spins in it have zero average value, see Eq. (28).

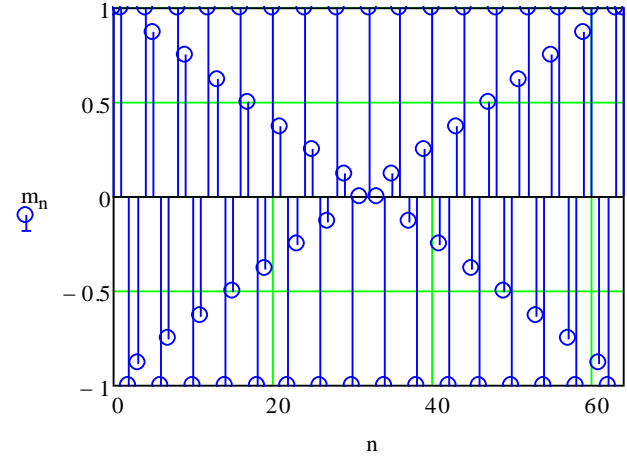

Fig.6. Magnetization profile at $\alpha < h_1 < 2\alpha - 1$ (F2 region) for $N = 65$.

Fourier transform of (26) is

$$\tilde{m}_k = \sum_{n=0}^{N-1} e^{ikn} m_n = 1 + 1/\cos(k), \quad k = 2\pi r / N = \pi r / 2M, \quad r = 0, 1, ..., N-1,$$

It also shows independence of its form on $N$ and singularities at $k = \pi/2$ and $k = 3\pi/2$.

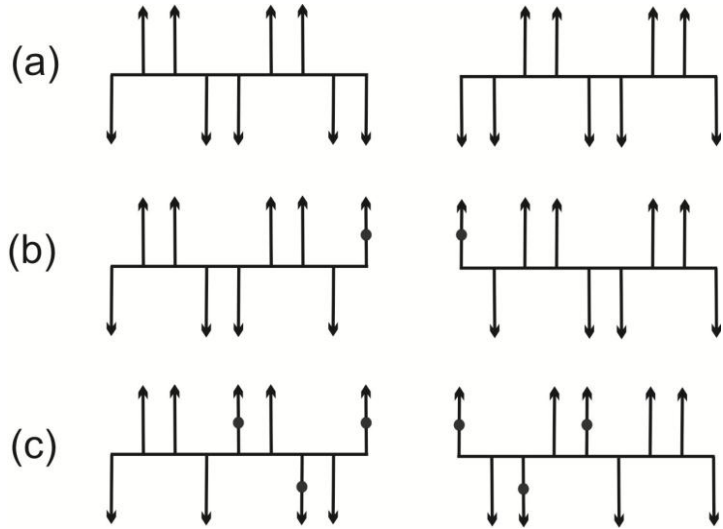

Fig.7. Procedure of ground state formation at $\alpha < 1$, $h_1 < 2\alpha - 1$ (F3 region) for $N = 9$. (a) Two parent states, below there are ground states obtained from them via flipping of one spin (b) and three spins (c). Flipped spins are marked by the dots.

The profile (27) at $\alpha < 1$, $h_1 < 2\alpha - 1$ (*F3* region) is the result of the averaging over $(N-1)/2 = 2M$ ground states. The procedure of their formation is shown in Fig. 7 for *N* = 9. It makes the number of the satisfied NN bonds larger at the expense of the NNN ones. This provides the energy gain as $\alpha < 1$.

## 6. Discussion and conclusions

Above results are summarized on the phase diagram in Fig. 2. *F1* phase exists in even chains and is quite similar to that in ordinary NN AF chain. In *F2* region the chains with *N* = 4*M* and *N* = 4*M* + 1 has modulated phases. *F3* phase of *N* = 4*M* + 1 chain is formally periodic (of $(0,0,+,0)$ type, cf. Eq. (27)) but it is also the result of the averaging over inhomogeneous kink states. Their number is always of order *N* due to the degeneracy with respect to the kink position. In the above examples the form of kink states can be readily guessed from the results of TM calculations but in more complex cases it can be also calculated via adding local fields lifting their degeneracy.

The mechanism of appearance of field-induced inhomogeneous ground states in finite AF chains lies in the degeneracy of their (infinite) AF ground states in a field. In the above examples the kink states are formed via conjunction of two shifted AF states which results in appearance of nonzero magnetic moment along the field. The energy gain from this may overcome the energy needed for the kink formation at large enough *H*. One may be interested if this mechanism can work in higher dimensions, non-Ising systems or quantum models having degenerate ground states. Now there is no definite answer to this, yet it seems worthwhile to study the finite-size effects in such systems.